\documentclass[journal, comsoc, final, twoside]{IEEEtran}

\usepackage[T1]{fontenc}

\usepackage[pdftex]{graphicx}
\usepackage{xcolor}
\usepackage[hidelinks]{hyperref}
\usepackage{url}

\usepackage{amsmath,amssymb,amsfonts}
\usepackage[cmintegrals]{newtxmath}
\usepackage{bm}
\usepackage{numprint}

\usepackage[super]{nth}

\usepackage{booktabs}

\usepackage{pgfplots}
\pgfplotsset{compat=newest}
\usepgfplotslibrary{groupplots}
\usepackage{tikz}
\usetikzlibrary{matrix, positioning, patterns, shapes, arrows}

\definecolor{Paired-2}{RGB}{166,206,227}
\definecolor{Paired-1}{RGB}{31,120,180}
\definecolor{Paired-4}{RGB}{178,223,138}
\definecolor{Paired-3}{RGB}{51,160,44}
\definecolor{Paired-6}{RGB}{251,154,153}
\definecolor{Paired-5}{RGB}{227,26,28}
\definecolor{Paired-8}{RGB}{253,191,111}
\definecolor{Paired-7}{RGB}{255,127,0}
\definecolor{Paired-10}{RGB}{202,178,214}
\definecolor{Paired-9}{RGB}{106,61,154}
\definecolor{Paired-12}{RGB}{255,255,153}
\definecolor{Paired-11}{RGB}{177,89,40}

\usepackage[ruled, linesnumbered, vlined]{algorithm2e}
\SetAlCapFnt{\footnotesize}
\SetAlFnt{\footnotesize}
\SetAlCapNameFnt{\footnotesize}

\SetCommentSty{mycommfont}

\usepackage{lipsum}

\usepackage[inline]{enumitem}
\usepackage{dirtytalk}

\usepackage{microtype}
\begin{document}
\bstctlcite{IEEEexample:BSTcontrol}
\title{On Systematic Polarization-Adjusted \\Convolutional (PAC) Codes}


\author{Thibaud~Tonnellier~and~Warren~J.~Gross,~\IEEEmembership{Senior~Member,~IEEE}%
\thanks{Manuscript received Month XX, 2020. The associate editor coordinating the
review of this letter and approving it for publication was X. XXX.}%
\thanks{T. Tonnellier and W.~J.~Gross are with the Department
of Electrical and Computer Engineering, McGill University, Montréal, Québec, Canada.}%
\thanks{e-mails: thibaud.tonnellier@mail.mcgill.ca,  warren.gross@mcgill.ca.}}


\maketitle
 
\begin{abstract}
Polarization-adjusted convolutional (PAC) codes were recently proposed and arouse the interest of the channel coding 
community because they were shown to approach theoretical bounds for the (128,64) code size.
In this letter, we propose systematic PAC codes. Thanks to the systematic property, improvement in the bit-error rate of up
to $0.2$ dB is observed, while preserving the frame-error rate performance. Moreover, a genetic-algorithm-based construction method 
targeted to approach the theoretical bound is provided. It is then shown that using the proposed construction
method systematic and non-systematic PAC codes can approach the theoretical bound even for higher code sizes such as (256,128).
\end{abstract}

\section{Introduction}
\IEEEPARstart{T}{he} last half-century has witnessed considerable advances in the field of channel coding. While at the 
beginning channel codes were often short due to complexity considerations, the development of iterative decoding 
algorithms have enabled longer codes to close the gap from Shannon limit to a few tenths of a decibel 
\cite{Berrou1993}. However, applications requiring short-to-moderate blocklength codes are emerging and the 
design of modern codes for such cases is challenging \cite{liva2016code}.

The well-known formula giving the capacity is only valid for the infinite
blocklength regime. To evaluate the quality of short-to-moderate length codes,
recent works provided refined analysis and derivations to estimate coding bounds in the finite-blocklength regime 
\cite{Polyanskiy10Channel, Erseghe16Finite}. 
In the remainder of this letter, we consider the \emph{normal approximation} (NA), which is a valid approximation
for both achievability and converse bounds. For more details, we refer the reader to \cite{Erseghe16Finite}. 

While polar codes can achieve the capacity of binary-input memoryless channels for infinite blocklength \cite{Arikan2009}, 
their finite-length decoding performance was rather poor. Substantial amount of work has been carried out to improve 
their performance. Among the various proposals, one of the major advances was the concatenation of a 
polar code with an outer CRC code \cite{Niu12CRC}. Using a list decoder, significant gains were observed, which accelerated 
the interest of industry and resulted in the standardization of polar codes in the 3GPP \nth{5} generation mobile network \cite{3GPP}. 
In \cite{Wang16Parity, Zhang18Parity}, some frozen bits are replaced by the output of a parity-check constraint. 
In \cite{Trifonov13, Trifonov16sub}, an extension of polar codes is proposed by considering dynamic frozen bits obtained 
as subcodes of eBCH codes. In \cite{Fazeli19Conv}, the conventional CRC code is replaced by a high-rate punctured convolutional code.

In \cite{arkan2019sequential}, polarization-adjusted convolutional (PAC) are proposed. They rely on the concatenation 
of a rate-1 convolutional code with a polar code. It was shown in \cite{arkan2019sequential} that a (128, 64) PAC code,
built around a convolutional code of constraint length 7 can meet the NA when decoded with the 
Fano algorithm. In \cite{cocskun2019efficient}, authors showed that the NA was also met using either a tail-biting 
convolutional code of constraint length 15 decoded with the wrap-around Viterbi algorithm (WAVA) \cite{Fossorier03}, 
or an eBCH code decoded with the ordered statistics decoding (OSD) algorithm with order 4 \cite{Fossorier95}. 
However, these two decoders have a high computational complexity. The WAVA algorithm requires several rounds of 
the Viterbi algorithm, whose complexity is exponential with the constraint length of the code. 
Also, for OSD of order $i$, the number of test error patterns is given by $\sum_{w=0}^i {k \choose w} $, which can be excessive.
On the other hand, PAC codes can be decoded with conventional polar decoders
by considering dynamic constraints on frozen bits. Moreover, the Fano decoding algorithm, which enabled to meet the NA, 
can exhibit a moderate computational complexity in the low error-rate regime, making PAC codes appealing.

In this letter, we propose a method to construct systematic PAC codes. An algorithm to construct the frozen set of 
non-systematic and systematic PAC codes is also given. Simulation results show that using the proposed construction method and the Fano decoding
algorithm, PAC codes can reach the NA for a wide range of FERs, even for short-to-moderate sizes such as (256, 128) PAC code. 
To the best of our knowledge, such results were not claimed previously in the literature. Moreover, bit-error rate (BER) is
further improved with systematic PAC by up to $0.2$ dB compared with the non-systematic PAC codes. 
Finally, a modified encoding is proposed, reducing the average decoding complexity by $13\%$ while maintaining the 
error-rate performance.

\section{Background}
\subsection{Polar codes} \label{sec:back:polar}
Polar codes (PCs) are named after the principle of channel polarization. This phenomenon causes two copies of 
a channel to be transformed into two synthetic channels, such that one becomes 
upgraded and the other one becomes degraded. The generator matrix $\bm{G}$ for a polar code of length 
$N$ is obtained by computing the $n\textsuperscript{th}$ Kronecker product, denoted $\otimes$, of the polarizing kernel 
$\bm{F_2} = \big[ \begin{smallmatrix} 1\ 0 \\ 1\ 1 \end{smallmatrix} \big]$, where $n = \log_2N$: 
$\bm{G} = \bm{F_2}^{\otimes n}$. 
Note that $\bm{G}$ is an $N \times N$ matrix, which differs from the typical  $K \times N$ matrix of a linear block code. 
Thus, the encoding process for a $(N,K)$ PC comprises two steps. First, the $K$-length message $\bm{u}$ is extended with 
$N-K$ bits whose the values are known.
Then, the encoding is carried out through the matrix multiplication: $\bm{x} = \bm{u}_e\cdot\bm{G}$.
To choose the location of the $N-K$ bits constituting the frozen set $\mathcal{F}$, several methods have been proposed
such as the density-evolution under Gaussian approximation (GA) \cite{Trifonov2012} or the $\beta-$expansion \cite{He2017}. 
If $\mathcal{F}$ is set to the indices with lowest row weight in $\bm{G}$, the polar code reverts to a Reed-Muller (RM) code.
It is also possible to design $\mathcal{F}$ while targeting a specific decoder using a genetic algorithm \cite{Elkelesh2019}.

The primary algorithm used to decode polar codes is known as the successive cancellation (SC) decoder \cite{Arikan2009}.
The SC decoder can be visualized as a binary tree traversal with left-branch priority. 
The tree has a depth of $n+1$ and $N$ leaf nodes, which represent the estimated codeword $\hat{\bm{u}}$. 
Each stage contains $2^{n-s}$ nodes, where $s \in [0, n]$ indicates the stage number counting from the bottom of the tree. 
Each node $v$ contains $N_v = 2^s$ log-likelihood ratios (LLRs) and bit 
partial sums noted $\bm{\alpha_v}$ and $\bm{\beta_v}$, respectively. 
At each leaf node, a hard 
decision is made on the $\alpha$ value to determine the corresponding $\beta$ value, unless the node corresponds
to a frozen index, whereby $\beta$ is known. After returning from a right branch, the partial sums in the parent 
node are updated as in the encoder.

The SC decoding algorithm suffers from two main drawbacks: it exhibits poor error correction performance 
for short-to-moderate blocklengths, and its sequential nature induces high latency. To 
tackle the later, simplified SC (SSC) \cite{Alamdar2011} and fast simplified SC (FSSC) \cite{sarkis2014fast} were
proposed. These methods identify nodes that can be 
efficiently decoded without descending the tree further, which essentially prunes the decoding tree.
To improve error correction performance, the SC-List (SCL) algorithm was introduced in \cite{Chen2012, Tal2015}. 
By duplicating paths at each leaf node corresponding to an information bit, a list of codeword path candidates are considered.
The list of paths is managed by maintaining only $L$ best candidates throughout decoding. 

\subsection{Convolutional codes}
Convolutional codes were proposed by Elias in 1955 as an alternative to block codes in \cite{Elias55}. By then, the goal
was to develop a variable-length code. The output of a convolutional code depends on the current entry, 
but also on the past ones. Thus, each coded bit can be expressed as $x_i  = \sum\limits_{j=0}^{\nu} g_j\times u_{i-j}$, 
where $\bm{g}$ is the generator polynomial of degree $\nu$. $\nu + 1$ is often called the constraint length, and 
$2^\nu$ gives the number of possible states for the encoder.

Several algorithms can be considered to decode convolutional codes. The most commonly used is the Viterbi algorithm 
\cite{Viterbi67}, which is a maximum likelihood sequence estimator working on the trellis of the code. 
When the constraint length is large
it may be advantageous---for computational complexity concerns---to consider sub-optimal decoders.
Examples to this are the stack algorithm \cite{jelinek69fast} or the Fano algorithm \cite{Fano63}, 
both belonging to the family of sequential decoding algorithms and working on the tree representation of the code. 
The Fano algorithm is a depth-first search and articulated around a threshold. The decoder moves \emph{forward} as long as the metric
of the current path exceeds the current \emph{threshold}. If that is not the case, it moves \emph{backward} to find another branch 
meeting the threshold constraint. If no satisfactory branch can be found, the threshold is loosened, and the forward 
search is restarted. For further elaboration and description of this algorithm for convolutional codes, we refer the reader 
to \cite[p. 518]{moon2005error}.
Note that adaptations of the stack and Fano algorithms have been proposed to decode polar codes in \cite{niu2012stack, jeong2019sc}.

\subsection{PAC codes}
\begin{figure}
  \centering\includegraphics[width=.9\columnwidth]{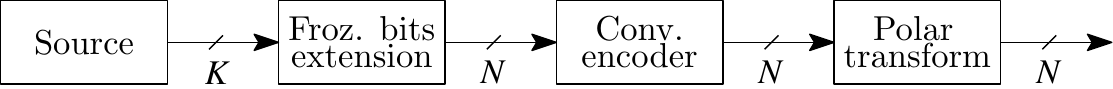}
  \caption{\label{fig:pac_enc}Encoding steps of a PAC code.}
\end{figure}
Polarization-adjusted convolutional (PAC) codes were recently proposed by Ar{\i}kan \cite{arkan2019sequential}. The 
principle of this new class of code is to concatenate an outer rate-1 convolutional code with the polar transform.
Thus, the encoding comprises 3 steps: first, the message is extended with frozen bits; then, the sequence is encoded 
using the convolutional code; and finally, the polarization matrix is used. Fig. \ref{fig:pac_enc} illustrates 
the encoding steps. The convolutional code enables a correlation
between the current bit and the previous bits. This correlation can be successfully exploited by a sequential decoding algorithm 
such as the Fano algorithm as reported in \cite{arkan2019sequential}.

Decoding PAC codes is similar to decoding convolutional codes when using the Fano algorithm.
The only differences are 
\begin{enumerate*}[label=(\arabic*)]
\item one branch only can stem at a node corresponding to a frozen location;
\item branch probabilities are not directly available at the channel output;
\item the threshold does not need to be tighten.
\end{enumerate*}
The metric used to rank list candidates during SCL decoding can be considered as the branch probabilities when transitioning
through the tree representation of the convolutional code. Thus, by only looking through the prism of the polar tree, 
PAC codes can be seen as polar codes with dynamic frozen bits \cite{Trifonov13}.
In detail, 
when a frozen location is reached, the input frozen value (usually 0) and the state 
of the convolutional code are used to obtain the parity and the future state of the convolutional code. The parity is 
then used in conjunction with the branch metric---the soft value obtained through the polar tree---to compute the path metric. 
When a location corresponding to a message bit is reached, the parity obtained as $\beta$ by traversing the polar tree is used 
to compute the future state of the convolutional code. 
Note that the value by which the threshold is increased in case no paths are found plays a crucial role. 
This value directly affects the speed and the decoding performance of the 
algorithm and is denoted by $\Delta$ in the following.
Finally, observe that by considering the simplified version of the path metric \cite{balatsoukas2015llr}, its value can 
only grow at a frozen location or when the ML decision is not considered (due to the need for a backward move). 
Therefore, a path change can only occur after the evaluation of a frozen location, simplifying the implementation of the
decoding algorithm. 

\subsection{Systematic codes}
An error-correcting code is said to be systematic if the message is explicitly found in the codeword. Systematic polar
codes have been proposed by Ar{\i}kan in \cite{arikan2011systematic}. Interestingly, systematic polar codes showed
an improved BER compared with their non-systematic counterparts; while the FER is 
similar \cite{arikan2011systematic}. In the following we consider the systematic polar encoder proposed 
in \cite{sarkis2014fast,sarkis2016flexible}. This
encoder can be seen as the successive application of two regular encodings with a re-freezing step in between.

The development of systematic convolutional codes is related to the discovery of turbo codes \cite{Berrou1993}. To achieve 
the performance of non-systematic convolutional codes with systematic codes, it is required to make them recursive: a 
combination of the encoder state is fed back to the input of the encoder. Such a code is called a recursive systematic 
convolutional code (RSCC).

\section{Systematic PAC codes}
In this Section, we propose systematic polarization-adjusted convolutional codes. Their encoding and construction are
first described. Then, a simplification reducing the computational complexity of the decoding processes is
proposed. 
\subsection{Construction of systematic PAC codes}
\begin{figure}[t]
  \setlength\abovecaptionskip{0\baselineskip} 
  \centering\includegraphics[width=.9\columnwidth]{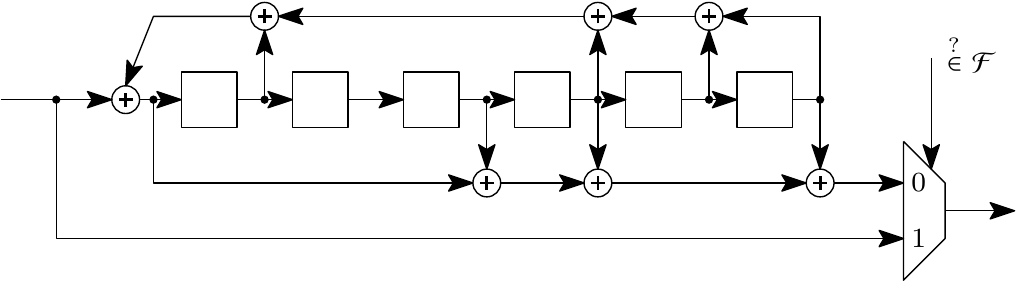}
  \caption{\label{fig:rscc_enc}Circuitry for the proposed rate-1 RSCC coder with generator polynomials $(115)_8$ and $(147)_8$.}
\end{figure}

The convolutional code considered in PAC codes is of rate 1, a rate that cannot be naturally achieved for an RSCC code. Indeed, 
the native rate of an RSCC encoder is at most 1/2 since one output is systematic, and the other is parity.
Hence, to obtain a rate of 1, we propose to connect a multiplexer to the two outputs of a rate 1/2 RSCC. Then, when the 
current location corresponds to a frozen index, the parity output is considered. Otherwise, the systematic output is 
considered. By doing so, from a vector of size N containing information and frozen bits, one can obtain a vector of 
size N whose frozen bits have been modified via the RSCC encoder. 
Fig. \ref{fig:rscc_enc} illustrates such an encoder with 
generator polynomial $(115)_8$ for the parity output and $(147)_8$ for the feedback connections.
Finding \say{good} polynomials is a complex problem---fortunately, due to the research effort following the discovery of 
turbo codes, lists of polynomials for RSCC codes are available in \cite{Benedetto98}.

Since, by definition, the proposed rate-1 RSCC encoder is systematic, the 
information bits are not modified. Hence, applying the RSCC coding step after the re-freezing step ensures both that 
the resulting sequence is systematic, and that the frozen bits are modified accordingly to the convolutional code. 
An algorithmic summary of the encoding steps is given in Algorithm \ref{alg:encoding_spac}. Note that, as stated in 
\cite{sarkis2016flexible}, constraints on the frozen set have to be ensured to make the encoding valid.

\begin{algorithm}[!t]
    \DontPrintSemicolon
    \caption{\label{alg:encoding_spac}Proposed systematic encoding of PAC code.}

    \KwIn{$\bm{u}$: word; $\mathcal{F}$: frozen set}
    \KwOut{$\bm{x}$: systematic codeword}

    \SetKwFunction{extendWithFB}{extendWithFrozenBits}
    \SetKwFunction{encodeRSCC}{encodeRSCC}
    \SetKwFunction{encodePolar}{encodePolar} 

    $\bm{u}_e \leftarrow $ \extendWithFB{$\bm{u}, \mathcal{F}$}\;
    $\bm{x} \leftarrow $ \encodePolar{$\bm{u}_e$}\tcp*[f]{$\bm{x}=\bm{x}\bm{G}$}\; 
    \ForEach{index $i$ in $\mathcal{F}$}
    {
      $x_i = 0$\; 
    }
    $\bm{x} \leftarrow $ \encodeRSCC{$\bm{x}$}\tcp*[f]{with e.g., Fig. \ref{fig:rscc_enc}} \;
    $\bm{x} \leftarrow $ \encodePolar{$\bm{x}$}\;
   
    \KwRet{$\bm{x}$}
\end{algorithm}

The frozen bit selection is an inherent research topic associated with polar codes. While there are analytical methods 
to obtain good frozen sets for regular polar codes decoded by the SC algorithm, none have been proposed for PAC codes.
In \cite{arkan2019sequential}, it was proposed to follow the RM rule to construct the frozen set. 
It is known that the 
RM construction results in codes with larger minimum distances than the ones obtained with other construction methods.
However, when the length and the rate of the code under 
consideration do not correspond to those of an RM code, this method cannot be directly applied \cite{li2014rmpolar}. 
We propose then the following method to efficiently construct non-systematic and systematic PAC codes. This technique is based upon the 
one proposed in \cite{Elkelesh2019}. However, instead of relying on the error rate performance as in in \cite{Elkelesh2019} 
for the fitness function, 
we propose to rely on the distance spectrum of the code. This method has the advantage of being independent of the noise 
realization and thus the undesired variations in the results are reduced. Moreover, our experiments demonstrated a reduced time complexity
since the computation of reliable FER at high SNR is more time-consuming than the estimation of the distance profile of the code.
Finally, since our objective with PAC codes is to be as close as possible to theoretical limits, it is necessary to 
improve the distance profiles, which can be exploited by the Fano algorithm.
Algorithm \ref{alg:construct} summarizes the proposed construction method. First an initial population of frozen sets
($\mathbb{P}$) is initialized using standard construction methods as 
presented in Section \ref{sec:back:polar} (line 1). 
The population is extended as proposed in \cite{Elkelesh2019} while ensuring all the frozen sets produce valid systematic
codes as aforementioned (line 3). Then, minimum distances are computed by using, for example, the efficient 
method proposed in \cite{Liu14}, based on the use of an SCL decoder with a really large list size. 
The frozen sets population is thereupon pruned by considering individuals with the best distance profile (lines 4-5).
The process is repeated until a maximum number of iterations is reached.

\begin{algorithm}[!t]
    \caption{\label{alg:construct}Modified genetic algorithm construction method.}
    \DontPrintSemicolon

    \SetKwData{P}{$\mathbb{P}$}
    \SetKwData{I}{$it_{max}$}

    \KwIn{$N, K,$ \I: maximum number of iterations}
    \KwOut{\P: population of frozen sets}

    \SetKwFunction{initializePopulation}{initializePopulation}
    \SetKwFunction{extendPopulation}{extendPopulation}
    \SetKwFunction{computeMinimumDistance}{computeMinimumDistance}
    \SetKwFunction{prunePopulationValidaty}{pruneInvalidCandidate}
    \SetKwFunction{prunePopulationFitness}{prunePopulation}

    \P $\leftarrow$ \initializePopulation{$N,K$} \;
    \For{$it=1,2,\ldots,\I$}{
      \P $\leftarrow$ \extendPopulation{$\P$} \;
      \P $\leftarrow$ \computeMinimumDistance{\P} \;
      \P $\leftarrow$ \prunePopulationFitness{\P} \;
    }
    \KwRet{\P}
\end{algorithm}

\subsection{Simplification of the decoding}
SSC decoding \cite{Alamdar2011} aims to stay at the top of the polar decoding 
tree as much as possible. This is realized by the identification of specific nodes in the decoding tree that do
not require to be explicitly traversed. Since the decoding of PAC codes still relies on the traversal of the polar decoding
tree, all the specific patterns already discovered for regular polar codes can be applied to PAC decoding.
However, in the case of PAC codes, the convolutional code is located at the bottom of the polar tree. It is, therefore, 
necessary to \say{un-polarize} the hard-decided partial sums obtained after each special node decoding, to update
the state of the convolutional code or to generate the dynamic frozen bits to be compared. Thus, due to the construction
of PAC codes, it is always necessary to go to the bottom of the polar tree. We now propose a solution to alleviate 
the need for the \say{un-polarize} operation after specific special nodes.

\begin{figure}[t]
  \setlength\abovecaptionskip{0\baselineskip} 
  \centering\includegraphics[width=.65\columnwidth]{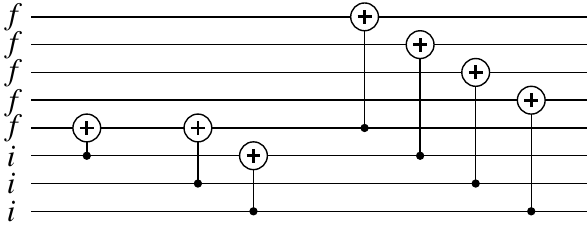}
  \caption{\label{fig:enc_simpl}Circuitry of the resulting encoding process, leading to a simplified decoding of PAC codes.
  $f$ and $i$ denotes respectively a frozen index and an information index.}
\end{figure}

A Rate-0 node is a node where all the indices are frozen. On the contrary, for a Rate-1 node, all the indices correspond
to information bits. Hence, since each of these nodes only involves bits of the same type, the polarization transform
applied on these nodes can be regarded as unnecessary. We, therefore, propose to remove the polarization transform for these 
special nodes. An example of the resulting encoder circuitry for a $PC(8,5)$ is given in Fig. \ref{fig:enc_simpl}. 
Observe that in this specific example, 5 XOR gates are removed. By removing these stages, the aforementioned steps at the 
decoding are also alleviated. Moreover, due to the scheduling of the Fano decoding algorithm that moves back and forth along 
the convolutional code tree, computations can be saved several times during the decoding of a frame. 
Finally, while un-polarization steps are not part of the critical path of polar decoder architectures 
\cite{Hashemi2017}, they would be part of it for a hypothetical hardware implementation of a PAC code decoder, 
because of the data dependency with the convolutional code. Removing these unnecessary steps is then even more appealing
even if minor pre-computations that only need to be performed once are added at the encoder side.

\section{Experimental Results}
In this section, simulation results of the proposed systematic PAC codes are reported. The error correction
performance is first considered and then followed by an evaluation of the decoding complexity reduction.
An AWGN channel and a binary phase-shift keying modulation are considered. A minimum of 200 frame errors is counted
for each SNR point to ensure accurate results. For a fair comparison, all the PAC codes use convolutional codes with a 
constraint length of 7. The polynomial for the non-systematic codes is $(131)_8$, which is usually considered as 
a \say{good} polynomial and already used in \cite{arkan2019sequential}. For the systematic codes, the
polynomials are $(115)_8$ for the parity output and $(147)_8$ for the feedback connections, corresponding to the encoder 
depicted in Fig. \ref{alg:encoding_spac}. All frozen sets are obtained via Algorithm \ref{alg:construct}. 
All the codes are decoded with the Fano algorithm. Rate-0 and Rate-1 nodes are not explicitly traversed and the parameter
$\Delta$ of the Fano algorithm is set to $2$ because our experiments showed a simulation speed-up with no impact on
the decoding performance compared to $\Delta= 1$.

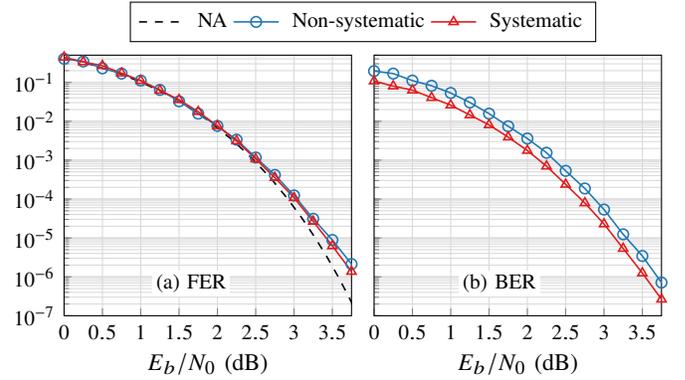
\begin{figure}[t]
\setlength\abovecaptionskip{-.1\baselineskip} 
    \begin{tikzpicture}
    \begin{groupplot}[group style={group name=12864, 
                                   group size= 2 by 1, 
                                   horizontal sep=3mm, 
                                   vertical sep=10mm}, 
                      footnotesize,
                      ymode=log,
                      grid=both, grid style={gray!30},
                      tick align=inside, tickpos=left,
                      xlabel=$E_b/N_0~\text{(dB)}$,
                      xlabel style={yshift=2pt},
                      xmin=0.0, xmax=3.75, xtick={0,0.5, 1,...,4},
                      /pgfplots/table/ignore chars={|},
                      ymax=0.5, ymin=1e-7,
                      ]

        \nextgroupplot[
                      width=.61\columnwidth,
                      height=.57\columnwidth,
                      legend style={font=\footnotesize, at={(1.04,1.05)}, anchor=south, legend columns=3},
                      legend cell align={left},
                      legend entries={~{NA}, ~{Non-systematic}, ~{Systematic}}]
            \addplot[mark=none    , black   , semithick, dashed]  table[x=Eb/N0, y=FER] {data/128_64_normal_approx.txt};
            \addplot[mark=o       , Paired-1, semithick        ]  table[x=Eb/N0, y=FER] {datav3/128_ns}; 
            \addplot[mark=triangle, Paired-5, semithick        ]  table[x=Eb/N0, y=FER] {datav3/128_s};

        \nextgroupplot[
                      width=.61\columnwidth,
                      height=.57\columnwidth,
                      yticklabels={,,}]
            \addplot[mark=o       , Paired-1, semithick        ]  table[x=Eb/N0, y=BER] {datav3/128_ns}; 
            \addplot[mark=triangle, Paired-5, semithick        ]  table[x=Eb/N0, y=BER] {datav3/128_s}; 
    
    \end{groupplot}
    \node[fill = white, draw=none, inner sep=0.5mm, above left = 0.25cm and -0.3cm of 12864 c1r1.south] {\footnotesize (a) FER};
    \node[fill = white, draw=none, inner sep=0.5mm, above left = 0.25cm and -0.3cm of 12864 c2r1.south] {\footnotesize (b) BER};
    \end{tikzpicture}
    \caption{\label{fig:128} PAC$(128, 64)$: systematic vs non-systematic.}
\end{figure}

Fig. \ref{fig:128} compares the decoding performance of the non-systematic PAC with its systematic counterpart for $N=128$
and $K=64$. In Fig. 
\ref{fig:128}(a), the FER is considered. The normal approximation curve is obtained via \cite{collins2016spectre}.
We observe that up to an FER of $10^{-3}$, all three curves are superimposed. At lower FERs, both PAC codes start 
diverging from the NA. For the highest SNR values plotted, systematic PAC has improved performance, but it is minimal. 
This behaviour can be explained by the slightly better distance profile of the systematic code as
reported in Table \ref{tab:distance}. Precisely, the multiplicity ($A_d$) associated with the distance of $16$
is $\numprint{2904}$ and $658$ for the non-systematic code and the systematic code, respectively.
Fig. \ref{fig:128}(b) compares the BER performance of the two codes. As expected, and compliant with \cite{arikan2011systematic},
a gain of approximatively $0.2$ dB is observed for the considered SNRs.

Fig. \ref{fig:256} plots the error-rate performance for $N=256$ and $K=128$. In Fig. \ref{fig:256}(a) one can observe 
that the NA is closely followed up to an FER of $10^{-5}$. To the best of the authors' knowledge, such a 
decoding performance was never reported in the literature for this size and code rate. Indeed, the best codes reported 
in \cite[Fig. 12]{cocskun2019efficient} are at least $0.4$ dB away from the NA.
The good distance profile obtained via the proposed modified genetic algorithm seems to be fully exploited by the
Fano algorithm and enables the observation of unmatched performance both for the non-systematic and systematic codes. 
Finally, regarding the BER performance portrayed
in Fig. \ref{fig:256}(b), the systematic code allows a gain of at least $0.1\text{ dB}$ more on the excellent performance of 
the non-systematic code. This gain even reaches $0.2\text{ dB}$ for FERs larger than $10^{-3}$.

To evaluate the complexity reduction, the number of steps required to decode frames was recorded during the decoding.
A step is defined as moving on the polar decoding tree from one layer to another or to the decoding of a special node
(i.e., assuming that enough computational resources are always available). Thus, for example, descending one layer, 
ascending one layer on the tree, and estimating the partial sums at a special node while updating the convolutional code state
are all considered as one decoding step. 
Due to the variable
latency property of the Fano decoding algorithm, the savings are not constant and depend on the target SNR. For the 
systematic (128,64) PAC code, we reported gains in the range $12\% - 14\%$ for $E_b/N_0 \in [0,3.5]$ dB. The gains 
grow with the SNR following a sigmoid-shaped function. Thus, by only removing unnecessary operations, the number of 
decoding steps is reduced by up to $14\%$, without incurring any decoding performance loss.


      

\section{Conclusion}
In this letter, we have presented a method to construct systematic PAC codes. 
A technique to design frozen sets for PAC codes was also proposed. The proposed construction method showed that the 
normal approximation bound can be closely approached for a code length of up to $256$ and rate one half. 
Due to the systematic property of the proposed codes, gains of up to $0.2$ dB 
for the BER were observed compared to their non-systematic counterparts, irrespective of the target error-rate. 
Finally, a simplification leading to a reduced decoding computational complexity of $~13\%$ was presented. By further 
improving the error-rate performance of PAC codes, this work contributes to the design of theoretical bound-approaching
codes in the short blocklength regime.

\begin{table}[t]
\setlength\abovecaptionskip{-.1\baselineskip} 
\centering
\caption{\label{tab:distance}First terms of the distance profile for the different codes considered, obtained via \cite{Liu14} with a list size of $\numprint{262144}$.}
\resizebox{\columnwidth}{!}{%
\begin{tabular}{rrl}
\toprule
Type &    $(N,K)$            & $d(A_d)$ \tabularnewline
\cmidrule(r){1-1}\cmidrule(r){2-2}\cmidrule(l){3-3}
  NS & $(128,\phantom{1}64)$ & $16(2904), 18(1916), 20(95776), 22(161547)$ \\
   S & $(128,\phantom{1}64)$ & $\begin{array}[c]{@{}l@{}}16(658), 18(559), 20(29936), 22(37739), \\24(44119), 26(66266), 28(82866)\end{array}$ \\
  NS & $(256,128)$           & $\begin{array}[c]{@{}l@{}}20(430), 22(68), 24(6709), 26(7), 28(14957), 30(136), \\32(105046), 34(143), 36(16342), 38(104), 40(50915)\end{array}$ \\
   S & $(256,128)$           & $\begin{array}[c]{@{}l@{}}20(9), 22(10), 24(900), 26(181), 28(9340), 30(29), \\32(58461), 34(177), 36(113132), 38(286), 40(79618)\end{array}$\\
\bottomrule
\end{tabular}}
\end{table}
\bibliographystyle{IEEEtran}
\bibliography{IEEEabrv,syst}

\begin{figure}[!t]
\setlength\abovecaptionskip{-.1\baselineskip} 
    \begin{tikzpicture}
    \begin{groupplot}[group style={group name=256128, 
                                   group size= 2 by 1, 
                                   horizontal sep=3mm, 
                                   vertical sep=10mm}, 
                      footnotesize,
                      ymode=log,
                      grid=both, grid style={gray!30},
                      tick align=inside, tickpos=left,
                      xlabel=$E_b/N_0~\text{(dB)}$,
                      xlabel style={yshift=2pt},
                      xmin=1, xmax=3, xtick={0,0.5,...,3},
                      /pgfplots/table/ignore chars={|},
                      ymax=0.1, ymin=1e-8,
                      ]

        \nextgroupplot[
                      width=.61\columnwidth,
                      height=.57\columnwidth,
                      legend style={font=\footnotesize, at={(1.04,1.05)}, anchor=south, legend columns=3},
                      legend cell align={left},
                      legend entries={~{NA}, ~{Non-systematic}, ~{Systematic}}]
            \addplot[mark=none    , black   , semithick, dashed]  table[x=Eb/N0, y=FER] {data/256_128_normal_approx.txt};
            \addplot[mark=o       , Paired-1, semithick        ]  table[x=Eb/N0, y=FER] {datav3/256_ns}; 
            \addplot[mark=triangle, Paired-5, semithick        ]  table[x=Eb/N0, y=FER] {datav3/256_s}; 
        
        \nextgroupplot[
                      width=.61\columnwidth,
                      height=.57\columnwidth,
                      yticklabels={,,}]
            \addplot[mark=o       , Paired-1, semithick        ]  table[x=Eb/N0, y=BER] {datav3/256_ns}; 
            \addplot[mark=triangle, Paired-5, semithick        ]  table[x=Eb/N0, y=BER] {datav3/256_s};
    
    \end{groupplot}
    \node[fill = white, draw=none, inner sep=0.5mm, above left = 0.25cm and -0.3cm of 256128 c1r1.south] {\footnotesize (a) FER};
    \node[fill = white, draw=none, inner sep=0.5mm, above left = 0.25cm and -0.3cm of 256128 c2r1.south] {\footnotesize (b) BER};
    \end{tikzpicture}
    \caption{\label{fig:256} PAC$(256, 128)$: systematic vs non-systematic.}
\end{figure}
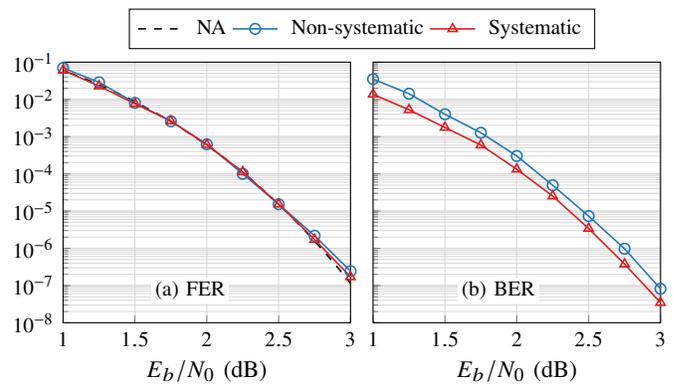

\end{document}